\documentclass[
    ,final            % use final for the camera ready runs
  ]
  {aipproc}

\usepackage{epsf}
\usepackage{graphicx}

\newcommand{\lQ}{\Lambda_{\rm QCD}}
\newcommand{\nn}{\nonumber}
\newcommand{\be}{\begin{equation}}
\newcommand{\ee}{\end{equation}}
\newcommand{\bea}{\begin{eqnarray}}
\newcommand{\eea}{\end{eqnarray}}

\newcommand{\als}{\alpha_{\rm s}}
\newcommand{\siml}{{\
\lower-1.2pt\vbox{\hbox{\rlap{$<$}\lower6pt\vbox{\hbox{$\sim$}}}}\ }}  
\newcommand{\simg}{{\
\lower-1.2pt\vbox{\hbox{\rlap{$>$}\lower6pt\vbox{\hbox{$\sim$}}}}\ }}  

\newcommand{\RS}{\rm RS}

\newcommand{\OS}{\rm OS}

\layoutstyle{6x9}

\begin{document}

\title{Hybrid potentials versus gluelumps}


\author{Antonio Pineda}{
  address={Grup de F\'\i sica 
Te\`orica and IFAE, Universitat Aut\`onoma de Barcelona, E-08193 Bellaterra, 
Barcelona, Spain}
}

%\author{<author2>}{
%  address={<common address for author2 and author3>}
%}

%\author{<author3>}{
%  address={<common address for author2 and author3>}
%  ,altaddress={<author1 address>} % additional visiting address
%}

\begin{abstract}
A potential model description of heavy quarkonium can be rigorously deduced from 
QCD under some circumstances. The potentials can be unambiguously 
related with Wilson loops with gluonic insertions, the spectral decomposition of which is 
a function of the spectrum and matrix elements solution of the static limit of NRQCD.
This spectrum is nothing but the static singlet potential and the hybrid potentials 
(which correspond to the gluonic excitations). We will quantitatively show that the latter 
unambiguously relate to the gluelumps at short distances using effective field theories.
\end{abstract}

\maketitle

%%%%%%%%%%%%%%%%%%%%%%%%%%%%%%%%%%%%%%%%%%%%
%% MAINMATTER
%%%%%%%%%%%%%%%%%%%%%%%%%%%%%%%%%%%%%%%%%%%%

\section{Introduction}

Potentials appear in a natural way in the description of systems with slow 
moving particles like in heavy quarkonium. It is now possible 
\cite{Brambilla:2000gk,Pineda:2000sz} to 
obtain a potential model-like description of the heavy quarkonium 
dynamics from first principles using effective field theories 
under some circumstances (for a review see \cite{Brambilla:2004jw}). 
The study of these 
potentials (the static \cite{Necco:2001xg}
and the relativistic \cite{Bali:1997am,Koma:2006si,Koma:2006fw} corrections) 
becomes then crucial as they dictate the dynamic of the
heavy quarkonium\footnote{One may wonder why bother to compute these potentials in the 
lattice, as one may try to do a direct computation of the heavy 
quarkonium properties in 
the lattice. Leaving aside the technical difficulties of 
these direct computations, in the best of the worlds they only provide 
with a very limited information on the properties of the heavy quarkonium: 
mainly a few masses (for the ground state and low excitations) and, maybe, some inclusive 
decays. On the other hand, with the potentials it is possible to 
obtain a detailed information of the shape of the heavy quarkonium, 
one can then compute the complete spectrum, all the inclusive 
decays and also opens the possibility to consider differential 
decay rates, which are far beyond the possibility of present 
direct lattice simulations.}. They can be obtained from the solution (spectrum 
and matrix elements) of NRQCD in the static limit. The 
spectrum of NRQCD directly provides the static singlet potential 
and the hybrid potentials. Therefore it is interesting to study them 
on their own. In particular, the study of the hybrid potentials may become 
important in order to discern whether the recently found resonances 
near threshold can be understood as hybrid states 
(see \cite{Eichten} for a review on the different possible interpretation
of these new states).

In this review we closely follow \cite{Bali:2003jq}. 
We will focus on the short distance limit of the hybrid potentials, 
specially on the aspects that can be fixed on symmetry arguments only, 
and we will quantitatively relate them with the physics of gluelumps.

\section{Static Limit of pNRQCD}

The static limit of NRQCD at short distances can be studied with the 
static version of pNRQCD. In this limit new symmetries arise. 

The pNRQCD Lagrangian at leading order in $1/m$ and in the multipole 
expansion reads~\cite{Pineda:1997bj,Brambilla:1999xf},
\begin{equation}
L_{\rm pNRQCD} =
\int d^3\!{\bf r}\,d^3{\bf R}\,{\rm Tr} \,\biggl[ {\rm S}^\dagger \left( i\partial_0 - V_s  \right) {\rm S} 
+ {\rm O}^\dagger \left( iD_0 - V_o \right) {\rm O} \biggr]
-\int d^3{\bf R} {1\over 4} F_{\mu \nu}^{a} F^{\mu \nu \, a}+O(r).
\label{pnrqcd0}
\end{equation}

All the gauge fields in Eq.\ (\ref {pnrqcd0}) are evaluated 
in ${\bf R}$ and $t$, in particular $F^{\mu \nu \, a} \equiv F^{\mu \nu \, a}
({\bf R},t)$ and $iD_0 {\rm O} \equiv i \partial_0 {\rm O} - g [A_0({\bf
  R},t),{\rm O}]$. 
The singlet and octet potentials
$V_{i}$, $i=s,o$ are to be regarded as matching coefficients, which depend
on the scale $\nu_{us}$ separating soft gluons from ultrasoft ones. In the
static limit ``soft'' energies are of $O(1/r)$ and ``ultrasoft'' energies
are of $O(\als/r)$. 
Notice that the hard scale, $m$, plays no r\^ole in this limit.
The only assumption 
made so far 
concerns the size of $r$, i.e.\ $1/r \gg \lQ$, such that the potentials can be 
computed in perturbation theory.  Also note that throughout this paper
we will adopt a Minkowski space-time notation.

The spectrum of the singlet state reads,
\be
\label{Es}
E_s(r)=2m_{\OS}+V_s(r)+O(r^2)
\,,
\ee
where $m_{\OS}$ denotes an on-shell (OS) mass.
One would normally apply pNRQCD to quarkonia and in this
case $m_{\OS}$ represents the heavy quark pole mass.
For the static hybrids, the spectrum reads
\be
\label{EH}
E_H(r)=2m_{\OS}+V_o(r)+\Lambda_H^{\OS}+O(r^2)
\,,
\ee
where
\be
\Lambda_H^{\OS}\equiv \lim_{T\to\infty} i\frac{\partial}{\partial T}\ln 
\langle H^a(T/2)\phi(T/2,-T/2)H^b(-T/2) \rangle
\label{LH}
\,.
\ee 
\bea
\phi(T/2,-T/2) &\equiv& \phi(T/2,{\bf R},-T/2,{\bf R}) \nn\\
&=& {\rm P} \exp \left\{ - ig \displaystyle \int_{-T/2}^{T/2} \!\! dt \, A_0({\bf R},t) \right\}
\,,
\eea
denotes the Schwinger line in the adjoint representation
and $H$ represents some gluonic field, for examples see
Table~\ref{gluedegen}.

Eq.~(\ref{EH}) allows us to relate the energies
of the static hybrids $E_H$ to the energies of the gluelumps,
 \be
\label{eqgl}
\Lambda_H^{\OS}=\left[E_H(r)-E_s(r)\right]-\left[V_o(r)-V_s(r)\right]+O(r^2).
\ee
This equation encapsulates one of the central ideas of this paper.
The combination $E_H-E_s$ is renormalon-free in perturbation theory
[up to possible $O(r^2)$ effects], and can be calculated unambiguously
non-perturbatively: the ultraviolet (UV) renormalons related to the infrared
(IR) renormalons
of twice the pole mass cancel each other. However, $\Lambda_H$
contains an UV renormalon that corresponds to the leading IR renormalon of
$V_o$.

\subsection{Symmetries of hybrid potentials and gluelumps}
The spectrum of open QCD string states can be completely classified by
the quantum numbers associated with the underlying symmetry group, up
to radial excitations.  In this case, these are the distance between the
endpoints, the gauge group representation under which these endpoints
transform (in what follows we consider the fundamental
representation), and the symmetry group of cylindrical rotations with
reflections $D_{\infty h}$. The irreducible representations of the
latter group are conventionally labelled by the spin along the axis,
$\Lambda$, where $\Sigma,\Pi,\Delta$ refer to $\Lambda=0,1,2$,
respectively, with a subscript $\eta=g$ for gerade (even) $PC=+$ or $\eta=u$
for ungerade (odd) $PC=-$ transformation properties.  All
$\Lambda\geq 1$ representations are two-dimensional. The
one-dimensional $\Sigma$ representations have, in addition to the
$\eta$ quantum number, a $\sigma_v$ parity with respect to
reflections on a plane that includes the two endpoints. This is
reflected in an additional $\pm$ superscript.  The state associated
with the static singlet potential transforms according to the
representation $\Sigma_g^+$ while the two lowest lying hybrid potentials
are within the $\Pi_u$ and $\Sigma_u^-$ representations, respectively.

In contrast, point-like QCD states are characterised by the $J^{PC}$
of the usual $O(3)\otimes {\mathcal C}$ rotation group as well as by
the gauge group representation of the source. In the pure gauge
sector, gauge invariance requires this representation to have
vanishing triality, such that the source can be screened to a singlet
by the glue. States created by operators in the singlet representation
are known as glueballs, octet states as gluelumps. In contrast to
gluelump states, where the octet source propagates through the gluonic
background, the normalization of glueball states with respect to the
vacuum energy is unambiguous.

Since $D_{\infty h}\subset
O(3)\otimes{\mathcal C}$, in the limit $r\rightarrow 0$ certain hybrid
levels must become degenerate. For instance, in this limit, the
$\Sigma_u^-$ state corresponds to a $J^{PC}=1^{+-}$ state with $J_z=0$ while
the $\Pi_u$ doublet corresponds to its $J_z=\pm 1$ partners.  The
gauge transformation property of the hybrid potential creation
operator will also change in this limit, ${\mathbf 3}\otimes{\mathbf
3^*}={\mathbf 1}\oplus{\mathbf 8}$, such that hybrids will either
approach gluelumps [cf.~Eq.~(\ref{EH})] or glueballs, in an
appropriate normalization. In the case of glueballs the correct
normalization can be obtained by considering the difference
$E_H(r)-E_s(r)$ from which the pole mass cancels. We will discuss the
situation with respect to gluelumps in detail below.

In perturbation theory, the ground state potential corresponds to
the singlet potential while hybrid potentials 
will have the perturbative expansion of the octet
potential, which have recently been computed to two loops \cite{Kniehl:2004rk}.

\subsection{Hybrid and gluelump mass splittings}
\label{sechsplit}
\begin{figure}
\epsfxsize=0.8\columnwidth
\includegraphics[width=0.85\columnwidth]{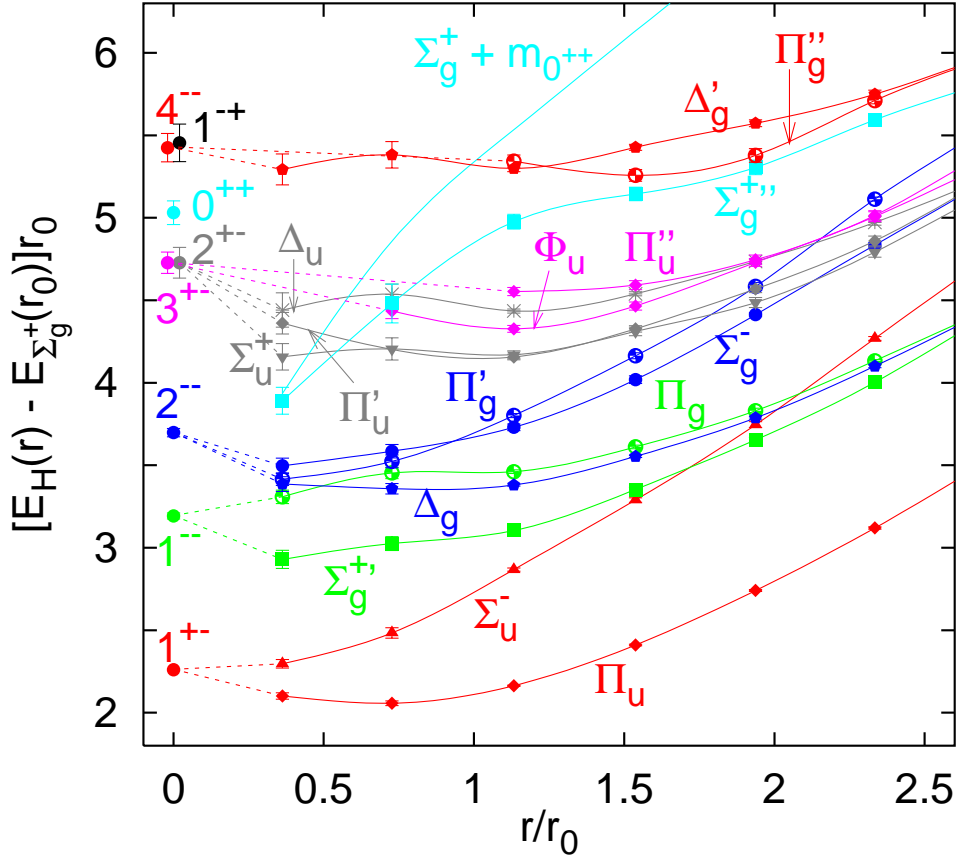}
\caption{\label{splitting}
Different hybrid potentials~\cite{Juge:2002br} at a
lattice spacing $a_{\sigma}\approx 0.2$~fm~$\approx 0.4\,r_0$,
where $r_0\approx 0.5$~fm, in
comparison with the gluelump spectrum, extrapolated to the continuum
limit~\cite{FM} (circles, left-most data points). The gluelump spectrum
has been shifted by an arbitrary constant to adjust the $1^{+-}$ state
with the $\Pi_u$ and $\Sigma_u^-$ potentials at short distance.  In
addition, we include the sum of the ground state ($\Sigma_g^+$)
potential and the scalar glueball mass $m_{0^{++}}$.
The lines are
drawn to guide the eye.}
\end{figure}

We would like to establish if lattice data on hybrid potentials
reproduces the degeneracies expected from the above discussion in the
short distance region. In the limit $r\rightarrow 0$, any given
$\Lambda\geq 1$ hybrid potential can be subduced from any $J^{PC}$ state with
$J\geq \Lambda$ and $PC=+$ for $\eta=g$ or $PC=-$ for $\eta=u$ representations.
For instance the $\Pi_u$ is embedded in
$1^{+-},1^{-+},2^{+-},2^{-+},\cdots$.  The situation is somewhat different
for $\Lambda=0$ states, which have the additional $\sigma_v$ parity:
the $\Sigma_g^+$ representation
can be obtained from $0^{++},1^{--},2^{++},\cdots$, $\Sigma_g^-$ from
$0^{--},1^{++},\cdots$, $\Sigma_u^+$ from $0^{+-},1^{-+},\cdots$ and
$\Sigma_u^-$ from $0^{-+},1^{+-},\cdots$.  We list all combinations of
interest to us in Table~\ref{gluedegen}.  The ordering of low lying
gluelumps has been established in Ref.~\cite{FM} and reads with
increasing mass: $1^{+-}, 1^{--}, 2^{--}, 2^{+-}, 3^{+-}, 0^{++},
4^{--}, 1^{-+}$,
with a $3^{--}$ state in the region of the $4^{--}$ and $1^{-+}$.
The $2^{+-}$ and $3^{+-}$ as well as the $4^{--}$ and
$1^{-+}$ states are degenerate within present statistical
uncertainties\footnote{The splittings between all states with respect
to the $1^{+-}$ ground state have been extrapolated 
to the continuum limit in
Ref.~\cite{FM} and we add our own
extrapolations for the $4^{--}$ and $1^{-+}$ states to these, based on
the tables of this reference.}.  The continuum limit gluelump masses
are displayed as circles at the left of Fig.~\ref{splitting}, where we
have added the (arbitrary) overall constant $2.26/r_0$ to the gluelump
splittings to match the hybrid potentials.  

%%%%%%%%%%%%%%%%%% Tabla gluedegen%%%%%%%%%%%%%%%%%%%%%%%%%%%%%%%%%%%%%%%5
\begin{table}
%\begin{ruledtabular}
\begin{tabular}{c|c|c|c|c}
point particle $J^{PC}$&$H$&$\Lambda_H^{\RS}r_0$&$\Lambda_H^{\RS}$/GeV&open string $\Lambda^{\sigma_v}_{\eta}$\\\hline
$1^{+-}$&$B_i$                           &2.25(39)&0.87(15)&$\Sigma_u^-, \Pi_u$\\
$1^{--}$&$E_i$                           &3.18(41)&1.25(16)&$\Sigma_g^{+\prime},\Pi_g$\\
$2^{--}$&$D_{\{i}B_{j\}}$                &3.69(42)&1.45(17)&$\Sigma_g^-,\Pi_g',\Delta_g$\\
$2^{+-}$&$D_{\{i}E_{j\}}$                &4.72(48)&1.86(19)&$\Sigma_u^+,\Pi_u',\Delta_u$\\
$3^{+-}$&$D_{\{i}D_{j}B_{k\}}$           &4.72(45)&1.86(18)&$\Sigma_u^{-'},\Pi_u'',\Delta_u',\Phi_u$\\
$0^{++}$&${\mathbf B}^2$                 &5.02(46)&1.98(18)&$\Sigma_g^{+\prime\prime}$\\
$4^{--}$&$D_{\{i}D_jD_kB_{l\}}$          &5.41(46)&2.13(18)&$\Sigma_g^{-\prime},\Pi_g'',\Delta_g',\Phi_g,\Gamma_g$\\
$1^{-+}$&$({\mathbf B}\wedge{\mathbf E})_i$&5.45(51)&2.15(20)&$\Sigma_u^{+\prime},\Pi_u'''$
\end{tabular}
%\end{ruledtabular}
\caption{\label{gluedegen}Expected degeneracies of hybrid potentials
at short distance, based on the level ordering of the gluelump spectrum.
Note that if the $3^{+-}$ gluelump turned out to be lighter than the $2^{+-}$
then the $\Sigma_u^{-'},\Pi_u',\Delta_u,\Phi_u$ potentials would approach the
$3^{+-}$ state while the $\Sigma_u^+,\Pi_u'',\Delta_u'$ potentials
would approach the $2^{+-}$ instead.}
\end{table}
%%%%%%%%%%%%%%%%%% End Tabla gluedegen%%%%%%%%%%%%%%%%%%%%%%%%%%%%%%%%%%%%%%%5

%%%%Figure splitting%%%%
%%%%End Figure splitting%%%%

Juge, Kuti and Morningstar~\cite{Juge:1997nc,Juge:2002br} have 
comprehensively determined the spectrum of hybrid
potentials. We convert their data,
computed at their smallest lattice spacing $a_{\sigma}\approx 0.2$~fm, 
into units of $r_0\approx 0.5$~fm.  Since the
results have been obtained with an improved action and on anisotropic
lattices with $a_{\tau}\approx a_{\sigma}/4$, one might expect lattice
artifacts to be small,
at least for the lower
lying potentials. Hence we compare these data, normalized to
$E_{\Sigma_g^+}(r_0)$, with the continuum expectations of the
gluelumps~\cite{FM}.  The full lines are cubic splines to guide the eye
while the dashed lines indicate the gluelumps towards which we would
expect the respective potentials to converge.

The first 7 hybrid potentials are compatible with the degeneracies
suggested by Table~\ref{gluedegen}.  The next state is trickier
since it is not clear whether $2^{+-}$ or $3^{+-}$ is lighter. In the
figure we depict the case for a light $2^{+-}$. This would mean that
$(\Sigma_u^+, \Pi_u',\Delta_u)$ approach the $2^{+-}$ while
$(\Sigma_u^{-\prime},\Pi_u'',\Delta_u',\Phi_u)$ approach the $3^{+-}$.
Note that of the latter four potentials only data for $\Pi_u''$ and
$\Phi_u$ are available. Also note that the continuum states
$\Pi_u',\Pi_u''$ and $\Phi_u$ are all obtained from the same $E_u$
lattice representation.  For the purpose of the figure we make an
arbitrary choice to distribute the former three
states among the $E_u',E_u''$ and
$E_u'''$ lattice potentials.  To firmly establish their ordering one
would have to investigate radial excitations in additional lattice
hybrid channels and/or clarify the gluelump spectrum in more detail.
Should the $2^{+-}$ and $3^{+-}$ hybrid levels be inverted then
$(\Sigma_u^{-\prime},\Pi_u',\Delta_u,\Phi_u)$ will converge to the
$3^{+-}$ while $(\Sigma_u^+,\Pi_u'',\Delta_u')$ will approach the
$2^{+-}$.  We note that the ordering of the hybrid potentials, with a
low $\Sigma_u^+$, makes the first interpretation more suggestive.

Finally the $\Sigma_g^{+\prime\prime}$ potential seems to head towards
the $0^{++}$ gluelump but suddenly turns downward, approaching the
(lighter) sum of ground state potential and scalar glueball 
instead.
The latter type of decay will eventually happen for all lattice
potentials but only at extremely short distances.  We also remark that
all potentials will diverge as $r\rightarrow 0$. This does not affect
our comparison with the gluelump results, since we have normalized
them to the $\Pi_u/\Sigma_u^-$ potentials at the shortest distance
available. (The gluelump values are plotted at $r=0$ to
simplify the figure.)

On a qualitative level the short-distance data are very consistent
with the expected degeneracies.  {}From the figure we see that at
$r\approx 2\,r_0\approx 1\,$fm the spectrum of hybrid potentials
displays the equi-distant band structure one would qualitatively expect
from a string picture. Clearly this region, as well as the cross-over
region to the short-distance behaviour $r_0<r< 2\, r_0$, cannot be
expected to be within the perturbative domain: at best one can possibly
imagine perturbation theory to be valid for the left-most 2 data
points. With the exception of the $\Pi_u$, $\Pi_u'$ and $\Phi_u$
potentials there are also no clear
signs for the onset of the short distance $1/r$ behaviour with a
positive coefficient as expected from perturbation
theory. Furthermore, most of the gaps within multiplets of hybrid
potentials, that are to leading order indicative of the size of the
non-perturbative $r^2$ term, are still quite significant, even at
$r=0.4\,r_0$; for instance the difference between the $\Sigma_u^-$ and
$\Pi_u$ potentials at this smallest distance is about 0.28~$r_0^{-1}\approx
110$~MeV.

\subsection{The difference between the $\Pi_u$ and $\Sigma_u^-$ hybrids}
\label{pisisplit}
\begin{figure}
\includegraphics[width=0.9\columnwidth]{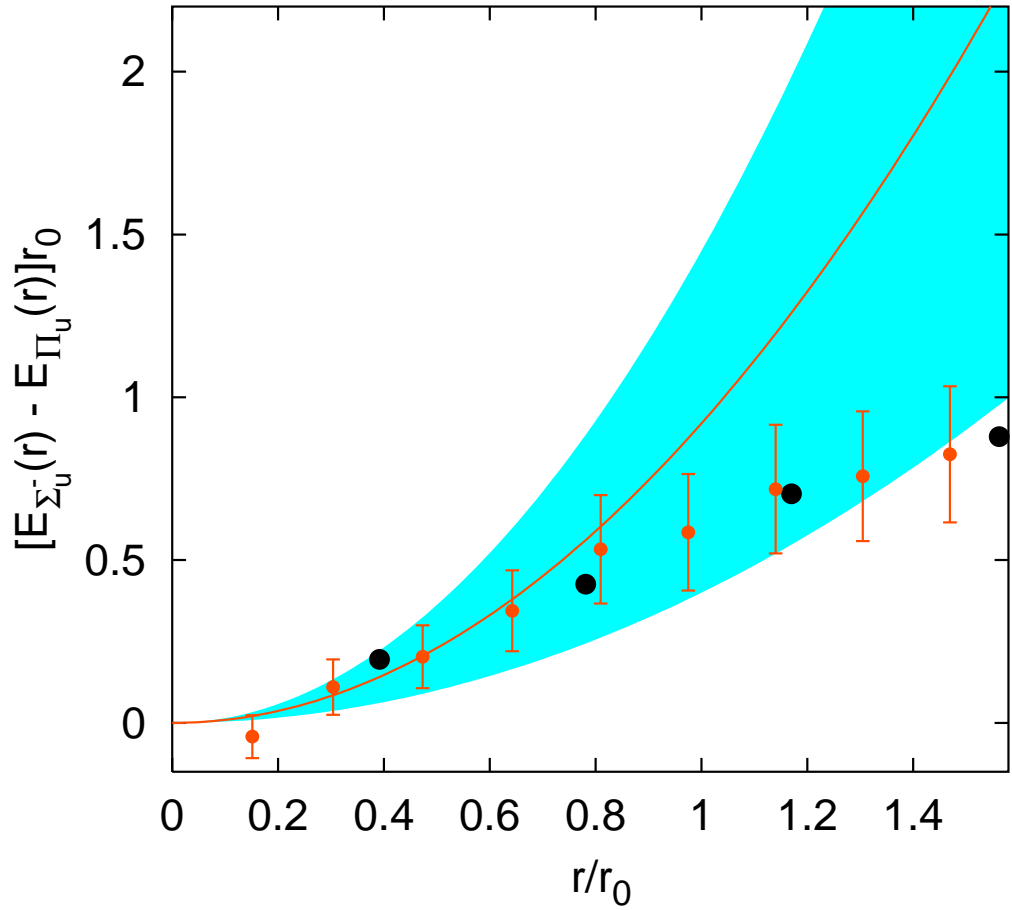}
\caption{\label{r2}
Splitting between the $\Sigma_u^-$ and the $\Pi_u$ 
potentials, extrapolated to the continuum limit,
and the comparison with a quadratic fit to the
$r\siml 0.5\,r_0$ data points ($r_0^{-1}\approx 0.4$~GeV). The big
circles correspond to the data of Juge et al.~\cite{Juge:2002br}, 
obtained at finite lattice spacing $a_{\sigma}\approx 0.39\,r_0$. 
The errors in this case are smaller than the symbols.}
\end{figure}

%%%%Figure r2%%%%
{}From the above considerations it is clear that
for a more quantitative study we need lattice data at
shorter distances, which have been obtained for the
lowest two gluonic excitations, $\Pi_u$ and $\Sigma_u^-$ in Ref. \cite{Bali:2003jq}. 
We display their
differences in the continuum limit in Fig.~\ref{r2}.
We see how these approach zero at small $r$, as expected
from the short distance expansion. pNRQCD predicts that the next effects should
be of $O(r^2)$ (and renormalon-free). In fact, we can fit
the lattice
data rather well with a $E_{\Pi_u}-E_{\Sigma_u^-} =A_{\Pi_u-\Sigma_u^-}r^2$ ansatz for
short distances, with slope
(see Fig. \ref{r2}),
\be
\label{eqa}
A_{\Pi_u-\Sigma_u^-}=0.92^{+0.53}_{-0.52}\,r_0^{-3}
\,,
\ee
where the error is purely statistical (lattice). This fit has been done using
points $r \siml 0.5\, r_0$. By increasing the fit range to $r \siml 0.8\, r_0$
the following result is obtained,
\be
A_{\Pi_u-\Sigma_u^-}=(0.83\pm 0.29)\,r_0^{-3}
\,,
\ee
indicating stability of the result Eq.~(\ref{eqa}) above.
%%%%End Figure r2%%%%

In order to estimate systematic errors one can add a quartic term: $b\,r^4$
(only even powers of $r$ appear in the multipole expansion of this quantity).
If the result is stable, our determination 
of $A_{\Pi_u-\Sigma_u^-}$ should not change much. Actually this is what
happens. If we fit up to $r\siml 0.5\, r_0$, we obtain
the central value $A_{\Pi_u-\Sigma_u}r_0^3=0.93$ with a very small quartic
coefficient,
$b\,r_0^5=-0.05$. If we increase the range to $r\siml 0.8\, r_0$,
we obtain the same central value,
$A_{\Pi_u-\Sigma_u^-}r_0^3=0.93$, but with a slightly bigger quartic term,
$b\,r_0^5=-0.18$. Introducing the quartic term enhances the stability
of $A_{\Pi_u-\Sigma_u^-}$ under variations of the fit range.
{}From this
discussion we conclude that the systematic error is negligible, in comparison
to the error displayed in our result Eq.~(\ref{eqa}).

We remark that within the
framework of static pNRQCD and to second order in the
multipole expansion, one can relate
the slope $A_{\Pi_u-\Sigma_u^-}$ to gluonic correlators of QCD.   

\section{Conclusions}

In conclusion, we have shown that the short distance behavior of the 
hybrid potentials is consistent with perturbation theory, the operator 
product expansion (effective field theories), and with the description 
of the leading non-perturbative effects in this regime in terms of the 
gluelump masses. For more details see \cite{Bali:2003jq}.

%\endinput
\end{document}